**MedSWFlow: An Open-Source LLM Workflow for Drafting Medical Social Work Case Plans**

Yulin Mao[1], Shiyu Li [1], Shuping Song [1], Yuling Zhang[1]

, Yajun Song[1*]

[1] School of Social and Public Administration, East China University of Science and Technology

* Corresponding author:

Yajun Song

School of Social and Public Administration
East China University of Science and Technology
Shanghai, China
Email: yj.song@ecust.edu.cn

**Acknowledgments**

We thank Professor Chenfeng Zhang for his technical advice and support in the development and testing of the AI-assisted workflow used in this study.

**Declaration of AI-assisted tool use**

The authors used ChatGPT (OpenAI) only for language editing, wording refinement, and manuscript formatting checks. All AI-assisted text was reviewed and revised by the authors, who take full responsibility for the final content of the article. ChatGPT was not listed as an author.

**Author Contributions**

**Yulin Mao** [ORCID: 0009-0001-7972-7266]**:** Development and refinement of the AI workflow, study implementation, interpretation of findings, manuscript reisivon, and manuscript review.

**Shiyu Li** [ORCID: 0009-0003-7127-6602]**:** Literature support, AI workflow co-design, Program Packaging, and manuscript review.

**Shuping Song** [ORCID: 0009-0007-2801-5369]**:** Literature review, AI workflow co-design, writing original draft, and manuscript review.

**Yuling Zhang** [ORCID: 0009-0003-5873-8518]**:** AI workflow co-design, and manuscript review.

**Yajun Song** [ORCID: 0000-0002-1543-4691]**:** Conceptualization, methodology, supervision, interpretation of findings, critical revision of the manuscript and correspondence with the journal.

All authors approved the final version of the manuscript and agreed to its submission.

**Statements and Declarations**

**Ethical Considerations**

The Research Ethics Committee of East China University of Science and Technology reviewed the study and confirmed that it was exempt from full ethical review **(No. 20260521051).** The case materials used in the study were de-identified and reconstructed before analysis and contained no directly identifiable information about patients, family members, or healthcare institutions. The expert review component involved professional evaluation of de-identified case materials and AI-generated draft case plans. It did not involve patient recruitment, clinical intervention, collection of sensitive identifiable information, or procedures beyond minimal risk.

**Consent to Participate**

The expert review component involved professional participants who evaluated de-identified case materials and AI-generated draft case plans. Participation was voluntary. The use of de-identified case materials and expert evaluation data was reviewed under the ethics exemption confirmed by the Research Ethics Committee of East China University of Science and Technology **(No. 20260521051).**

**Consent for Publication**

Not applicable.

**Declaration of Conflicting Interest**

The authors declared no potential conflicts of interest with respect to the research, authorship, and/or publication of this article.

**Funding Statement**

The preparation of this manuscript was supported by the Shanghai Philosophy and Social Science Planning Program, “Research on the Full-Cycle Linkage Mechanism between Medical and Community Health Social Work in Shanghai in the Context of Population Ageing” **[grant number 2025BSH007].** The funder had no role in the study design, data collection, analysis, interpretation of data, writing of the manuscript, or the decision to submit the article for publication.

**Data Availability Statement**

The data that support the findings of this study are not publicly available due to ethical and privacy restrictions. De-identified data may be made available from the corresponding author upon reasonable request and subject to institutional approval.

**MedSWFlow: An Open-Source LLM Workflow for Drafting Medical Social Work Case Plans**

**Abstract**

We present MedSWFlow, an open-source, model-agnostic LLM workflow for drafting medical social work case plans. The framework translates professional case-planning tasks into six stages: assessment, problem analysis, goal setting, intervention planning, risk anticipation, and planned effect evaluation. Drawing on established social work and behavioral frameworks, MedSWFlow standardizes case inputs, builds structured case profiles, and generates reviewable assessment forms and service plans through staged prompting. The system is released as an open-source research framework for reproducible case-plan generation across LLM providers. Outputs are intended as practitioner-reviewed drafts rather than final service decisions. Source code: https://github.com/santhiyacw-droid/MedSWFlow/tree/main.

## 1. Introduction

Large language models (LLMs) have made it easier to generate fluent professional text. In social work and human service settings, recent studies suggest that artificial intelligence may support tasks such as client identification, intervention classification, risk prevention, service monitoring, and information extraction from case narratives (Li et al., 2026; Perron, Luan, et al., 2025). Clinical social workers may also find value in LLMs for drafting and administrative tasks, while remaining concerned about confidentiality, empathy, contextual understanding, and loss of nuance (Báez et al., 2026). Responsible AI integration therefore requires attention not only to technical performance but also to professional oversight, task sensitivity, and potential effects on client wellbeing (Perron, Goldkind, et al., 2025).

Medical social work case planning is a demanding application because it requires more than coherent text. A case plan must transform referral information into assessment, problem prioritization, goals, interventions, risk considerations, and evaluation strategies. It must also account for illness conditions, family dynamics, caregiver capacity, service boundaries, resource availability, and institutional constraints. The person-in-environment perspective emphasizes understanding clients within their social and environmental contexts (Karls & Wandrei, 1992), while the biopsychosocial model views health outcomes as shaped by biological, psychological, and social factors (Engel, 1977). In hospital settings, medical social workers often address discharge planning, treatment communication, financial assistance, caregiver negotiation, crisis intervention, and interdisciplinary coordination under time pressure.

The limitations of fluent AI outputs are especially relevant in health and human service contexts. Research shows that patients may rate LLM-generated responses highly even when physicians judge them as less accurate or clinically useful than professional answers (Ye et al., 2024). In social work, AI systems can perform well on examination-style questions while still producing hallucinations, misunderstandings, or flawed reasoning (Zhao et al., 2025). AI-based predictions may also overlook contextual factors that are important for intervention planning (Lehtiniemi, 2024). These concerns do not preclude LLM use in case planning, but they suggest that outputs should be structured for practitioner review, revision, and contextual calibration.

A workflow-guided approach offers one way to make LLM-assisted planning more transparent. Rather than generating an entire case plan through a single prompt, a workflow divides planning into defined tasks, each with specified inputs, outputs, and constraints. This approach does not replace professional judgment. It makes planning logic more visible and supports comparison across connected LLM providers by applying the same case information and planning sequence. In high-risk human service settings, this structure can help practitioners identify unsupported assumptions, unrealistic recommendations, ethical concerns, and excessive service burden.

This preprint introduces MedSWFlow, an open-source, model-agnostic LLM workflow for drafting medical social work case plans. The workflow is the same six-step planning process used in our companion mixed-methods evaluation of AI-generated medical social work case plans (Song et al., 2026). The companion paper reports expert assessment of workflow-generated plans. This preprint focuses on the workflow itself, including its professional foundations, architecture, input structure, planning mechanism, safety constraints, implementation, and usage cases.

MedSWFlow translates medical social work planning into six stages: case assessment, problem analysis, goal setting, intervention planning, risk anticipation, and planned effect evaluation. The workflow draws on established frameworks, including the person-in-environment perspective, the biopsychosocial model, the International Classification of Functioning, Disability and Health (ICF), the COM-B model and Behaviour Change Wheel, goal-setting principles, intervention specification frameworks, and implementation outcome concepts (Bovend'Eerdt et al., 2009; Hoffmann et al., 2014; Michie et al., 2011; Proctor et al., 2011; Wade, 2009; World Health Organization, 2001). These frameworks are translated into workflow functions for organizing information, identifying priorities, specifying goals, designing interventions, anticipating risks, and defining evaluation indicators.

The contribution of this preprint is procedural and infrastructural. It describes how medical social work case planning can be represented as a staged workflow rather than a single prompt. It introduces a standardized case input structure and planning-oriented case profile. It explains how staged prompts generate draft assessment forms and service plans for practitioner review. It also identifies safety, assumption-control, and human-review requirements for planning involving vulnerable clients and health-related decisions.

MedSWFlow is not intended to replace medical social workers, clinical teams, or supervisory review. It does not provide medical diagnoses, treatment recommendations, legal determinations, or final service decisions. Its outputs are reviewable drafts that require verification of resources, feasibility, client preferences, privacy protections, and contextual appropriateness. The remainder of this paper introduces the theoretical foundations, system architecture, workflow design, safety considerations, implementation details, usage cases, limitations, and future work.

## 2. Professional and Theoretical Foundations

Medical social work case planning involves a sequence of tasks, including assessment, problem identification, goal setting, intervention design, risk anticipation, and outcome review. No single framework supports all of these activities. MedSWFlow therefore adopts a layered professional foundation. Broad social work and health perspectives guide how cases are represented, while behavioral, rehabilitation, intervention, and implementation frameworks support specific workflow stages.

Rather than serving only as background theory, these foundations are translated into system functions, including the case input structure, case profile generation, six-step workflow, prompt templates, and review constraints (Table 1).

**Table 1**
*Professional Foundations and Their Translation into MedSWFlow*

| Professional foundation | Planning role | System translation in MedSWFlow |
|---|---|---|
| Person-in-environment perspective (Karls & Wandrei, 1992) | Frames problems as interactions between individuals and their environments | Supports the five-module case input structure, including family context, resources, referral context, and environmental barriers |
| Biopsychosocial model (Engel, 1977) | Situates health within biological, psychological, and social conditions | Guides overall case representation and prevents reduction of cases to medical diagnoses alone |
| International Classification of Functioning, Disability and Health (World Health Organization, 2001) | Describes functioning, participation, and contextual factors | Structures the assessment stage and organizes referral information into an assessment draft |

| Professional foundation | Planning role | System translation in MedSWFlow |
|---|---|---|
| COM-B model and Behaviour Change Wheel (Michie et al., 2011) | Identifies barriers and facilitators through capability, opportunity, and motivation | Structures problem analysis and intervention direction identification |
| Goal-setting principles in rehabilitation (Bovend'Eerdt et al., 2009; Wade, 2009) | Converts problems into specific and measurable goals | Structures short-, medium-, and longer-term goal setting |
| Intervention specification frameworks, including fidelity, TIDieR, and AACTT (Borrelli, 2011; Hoffmann et al., 2014; Presseau et al., 2019; Proctor et al., 2013) | Specifies intervention content, actors, targets, and delivery conditions | Structures intervention planning and requires explicit action descriptions |
| Implementation outcomes and unintended consequences frameworks (Proctor et al., 2011; Stratil et al., 2024) | Highlights feasibility, burden, sustainability, and unintended effects | Structures risk anticipation and planned evaluation |
| Social work ethics and human oversight (Perron, Goldkind, et al., 2025; Reamer, 2023; World Health Organization, 2021) | Defines limits of AI-assisted planning | Supports privacy, consent, role-boundary constraints, and mandatory human review |

The person-in-environment perspective and biopsychosocial model provide the overall orientation. Medical social work considers illness together with family relationships, caregiving capacity, financial strain, emotional wellbeing, social roles, institutional factors, and service access. These perspectives are reflected in the five-module case input structure: client background and medical status, social and family context, psychosocial problems, resources and institutional constraints, and referral source and initial service request.

The assessment stage is informed by the ICF framework (World Health Organization, 2001). MedSWFlow uses ICF-inspired domains to organize information about functioning, participation, environmental influences, and personal circumstances. The system does not perform formal ICF coding. ICF is used as an organizing structure for transforming referral information into a draft assessment.

Problem analysis is guided by the COM-B model and Behaviour Change Wheel (Michie et al., 2011). Many medical social work interventions require change involving clients, caregivers, families, professionals, or institutions. Difficulties may arise from limitations in capability, opportunity, motivation, resources, or support. MedSWFlow uses this logic to analyze one or two priority problems and identify possible intervention directions.

Goal setting draws on rehabilitation goal-setting principles (Bovend'Eerdt et al., 2009; Wade, 2009). AI-generated goals can be broad or difficult to evaluate. MedSWFlow therefore requires goals to be linked to identified problems and expressed in observable, time-related terms. Depending on the case, goals may be organized into short-, medium-, and longer-term drafts for practitioner review.

Intervention planning is informed by intervention specification frameworks (Borrelli, 2011; Hoffmann et al., 2014; Presseau et al., 2019; Proctor et al., 2013). Rather than generating general recommendations, MedSWFlow requires proposed actions, responsible actors, targets, context, timing, intensity, and monitoring arrangements to be made explicit.

Risk anticipation and planned evaluation draw on implementation outcome and unintended consequence frameworks (Proctor et al., 2011; Stratil et al., 2024). Social work plans may create burden, rely on unavailable resources, or raise ethical concerns. MedSWFlow therefore includes dedicated stages for identifying implementation risks and defining outcome and process indicators.

Ethical and human oversight principles are operationalized through assumption control, role-boundary constraints, and human-review checkpoints. These safeguards define the limits of AI-assisted drafting and are detailed in Section 6.

**3. System Architecture**

MedSWFlow is a five-layer workflow system for generating reviewable medical social work case-plan drafts. The system is distributed as a local application that users run on their own device and connect to supported LLM providers using their own API credentials. MedSWFlow is therefore model-agnostic. It is not itself a language model. Its primary function is to control the professional workflow through which case information is standardized, profiled, transformed into staged prompts, and assembled into draft documents for practitioner review.

The architecture follows three principles. First, workflow logic is separated from model text generation. The workflow determines the planning sequence and output structure, while the connected LLM generates text within those constraints. Second, inputs and intermediate outputs are standardized to support reproducibility across connected LLM providers. Third, all generated content is treated as draft material for human review rather than autonomous decision-making.

Users configure model access, enter case information, and initiate workflow execution through the application. MedSWFlow then manages input normalization, case profiling, prompt construction, workflow execution, and document assembly while communicating with the selected LLM provider through the configured API connection.

A case moves through MedSWFlow in five steps (Table 2). First, the input normalization layer converts raw or structured case information into a standardized format. Second, the case profiling layer extracts planning-relevant information, including medical status, psychosocial concerns, family context, resources, constraints, referral information, and missing data. Third, the workflow layer processes the profile through six stages: assessment, problem analysis, goal setting, intervention planning, risk anticipation, and planned effect evaluation. Fourth, the prompt-generation layer converts each stage into structured prompts and sends them to the selected LLM provider. Fifth, the output layer assembles the generated content into reviewable documents, typically a case assessment form and a service plan. Figure 1 illustrates this end-to-end workflow from case input to the generation of reviewable documents.

**Table 2**

*Five-Layer Architecture of MedSWFlow*

| Layer | Main function | Input | Output |
|---|---|---|---|
| Input normalization layer | Standardizes raw or structured case information | Referral text, intake notes, structured case fields, or user-entered case description | Standardized case input |
| Case profiling layer | Extracts planning-relevant features and missing information | Standardized case input | Planning-oriented case profile |
| Six-step planning workflow layer | Controls the professional sequence of case planning | Case profile and previous workflow outputs | Staged outputs for assessment, analysis, goals, interventions, risks, and evaluation |
| Prompt-generation layer | Converts workflow tasks into structured prompts and communicates with the selected LLM provider | Workflow instructions, case profile, previous outputs, and API connection | Generated text for each workflow stage |
| Reviewable document output layer | Assembles generated outputs into practitioner-facing documents | Generated outputs and document template | Draft case assessment form and service plan |

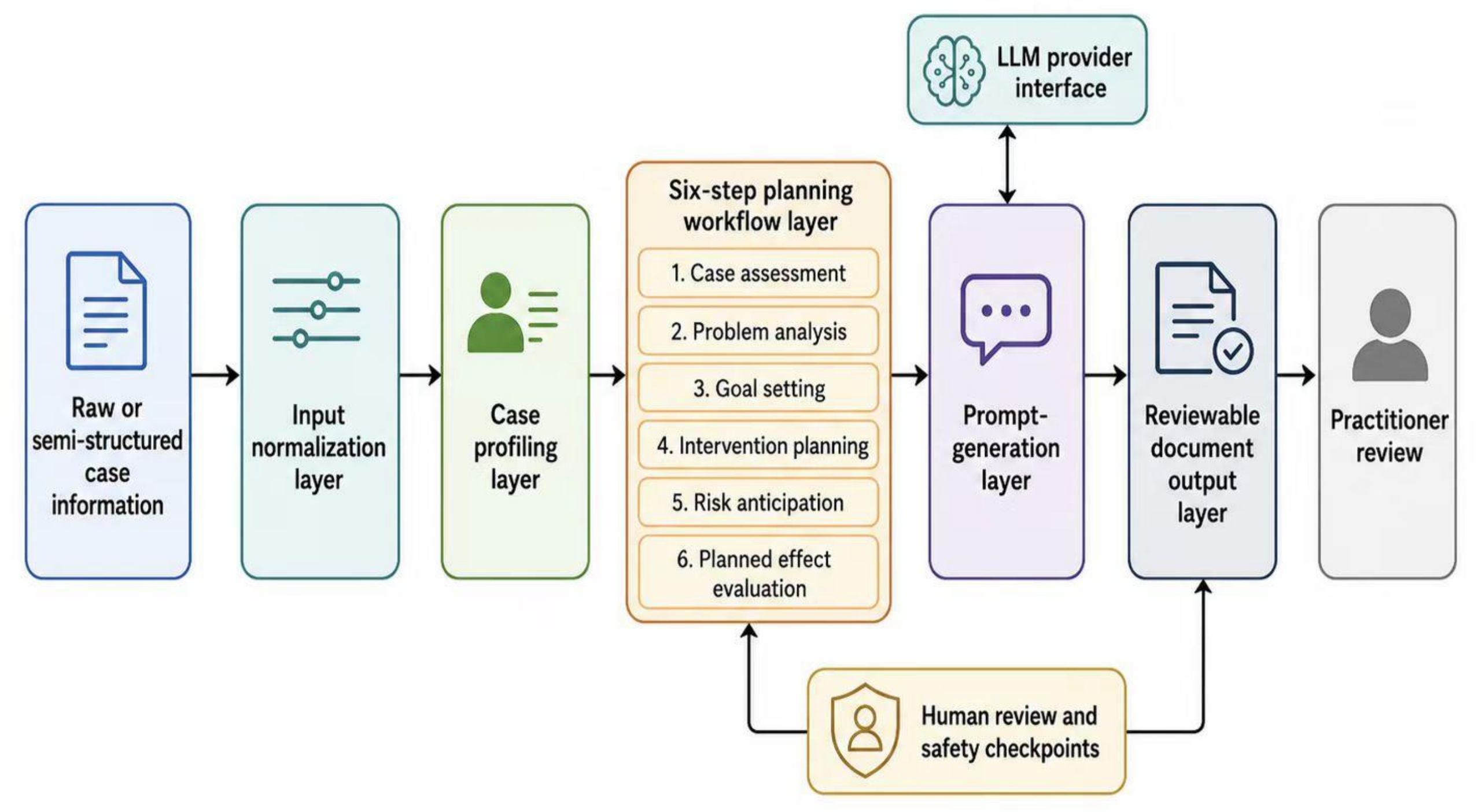

**Figure 1 MedSWFlow system architecture.** MedSWFlow organizes case planning into a five-layerworkflow. The connected LLM generates text within staged prompts but does not determine theprofessional sequence or replace practitioner review.

### 3.1 Input Normalization and Case Profiling

The input normalization layer prepares referral information for downstream processing. It organizes available information into a standardized structure aligned with medical social work practice and marks missing or unclear information rather than inferring it. The case profiling layer then converts the standardized input into a planning-oriented representation. This profile includes medical and functional status, psychosocial concerns, caregiving capacity, resource access, institutional context, ethical or relational flags, and information gaps. Section 4 describes this mechanism in greater detail.

### 3.2 Workflow and Prompt Generation

The workflow layer controls the six-step planning sequence. Each stage receives defined inputs and produces outputs that feed into later stages. The connected LLM generates text within this predefined structure rather than determining the sequence itself. The prompt-generation layer translates each workflow stage into structured instructions containing task definitions, relevant inputs, output requirements, and constraints. MedSWFlow does not require model training or fine-tuning. It relies on structured prompting and stage-by-stage information transfer.

### 3.3 Reviewable Document Output

The output layer assembles generated content into practitioner-facing draft documents. In the current design, two documents are produced: a case assessment form and a service plan. The assessment form contains assessment findings, problem analysis, and goals. The service plan contains intervention modules, anticipated risks, and evaluation indicators. These outputs do not authorize intervention, determine eligibility, replace consent procedures, or substitute for supervision. Their purpose is to provide a structured planning scaffold for practitioner review.

### 3.4 Auditability and Reproducibility

The five-layer architecture supports auditability and reproducibility by separating case inputs, case profiles, workflow stages, prompts, and final documents. This separation makes it easier to trace how specific planning statements were generated. The same workflow can also be executed across connected LLM providers through user-configured API connections. This allows comparison under a shared planning structure while keeping final judgment under human responsibility.

## 4. Case Input and Case Profile

MedSWFlow begins with case input standardization. Medical social work plans are sensitive to how referral information is described. Free-text referrals may emphasize diagnosis, discharge pressure, family conflict, emotional distress, financial strain, or institutional barriers. Without structure, an LLM may overemphasize salient details and overlook less visible but practice-relevant factors. MedSWFlow therefore separates case input from case planning. Available information is first organized into a standardized structure and then transformed into a planning-oriented case profile.

The case input schema is grounded in the person-in-environment perspective and the biopsychosocial model (Engel, 1977; Karls & Wandrei, 1992). The schema is organized into five modules, each with a specific focus (Table 3).It focuses on information commonly available at referral or intake and sufficient to support an initial draft plan.

**Table 3**

*Five-Module Case Input Structure*

| Module | Description | Example information |
|---|---|---|
| Client background and medical status | Demographics, diagnosis, treatment stage, functional status, and clinical background | Age, diagnosis, mobility limitations, pain, cognitive status, discharge pressure |
| Social and family context | Family structure, caregiving arrangements, social relationships, and support systems | Main caregiver, family conflict, caregiving burden, social isolation |
| Psychosocial problems | Emotional, behavioral, relational, or social concerns requiring intervention | Anxiety, grief, self-harm risk, treatment refusal, caregiver distress |

| Module | Description | Example information |
|---|---|---|
| Resources and institutional constraints | Resources, service access, financial strain, policy barriers, and institutional conditions | Insurance limits, housing crisis, charity support, community services |
| Referral source and initial service request | Referral pathway and requested social work involvement | Referral from physician or nurse; request for discharge planning or resource linkage |

These modules support later workflow stages. Medical information informs assessment. Psychosocial and family information supports problem analysis. Resources and institutional constraints guide intervention planning and risk anticipation. Referral information clarifies service expectations and role boundaries. The structure does not guarantee completeness, but it makes available and missing information more visible.

MedSWFlow supports two input modes. In guided structured input, users enter information directly into the five modules. In free-text input, users provide a narrative referral, intake note, or case summary, which MedSWFlow reorganizes into the same structure. Both modes produce a standardized format for downstream planning.

Information completeness checking is integrated into the input stage. Rather than relying only on a field checklist, MedSWFlow evaluates whether the case contains enough information to connect presenting problems, social work involvement, and planning needs. Three categories of incomplete information are distinguished: key missing information, information needing clarification, and recommended supplementary information. Key missing information refers to gaps that may disrupt planning logic. Information needing clarification includes vague or inconsistent statements. Recommended supplementary information includes contextual details that may strengthen planning but are not essential for a basic draft.

After normalization and completeness checking, MedSWFlow generates a case planning profile. This profile is not a diagnosis or professional assessment. It reorganizes case information into fields required by the six-step workflow (Table 4). This content is intended only for intermediate system storage and will not be included in the final case plan.

**Table 4**

*Case Planning Profile Fields and Planning Use*

| Profile field | Primary source | Planning use |
|---|---|---|
| Medical and functional status | Client background and | Supports assessment, functional needs |

| Profile field | Primary source | Planning use |
|---|---|---|
| | medical status | identification, and risk awareness |
| Psychosocial concerns | Psychosocial problems and referral request | Supports problem prioritization and goal formulation |
| Family and caregiving capacity | Social and family context | Supports analysis of caregiver burden, family roles, and intervention targets |
| Resource access and constraints | Resources and institutional constraints | Supports feasibility review and service linkage planning |
| Institutional and referral context | Referral source and service request | Supports role clarification and alignment with hospital workflow |
| Ethical and relational flags | Psychosocial, family, and referral information | Supports review of autonomy, consent, privacy, and trust issues |
| Missing information and assumptions to be checked | Completeness checking output | Supports assumption control and practitioner follow-up |

The distinction between case input and case profile is important. The input schema organizes information as provided by the user, whereas the profile reorganizes it for planning purposes. For example, a family member refusing discharge may appear under family context but later be represented as caregiving capacity, family conflict, institutional pressure, and an ethical concern. This layered representation allows the same information to support multiple planning tasks.

Assumption control is a central design feature. Unsupported information must be marked as unknown or requiring further assessment. The workflow should not invent family resources, client preferences, community services, medical facts, or institutional permissions. For lengthy case descriptions, MedSWFlow may summarize information before downstream processing. Summaries should preserve planning-relevant details, including medical facts, social work triggers, family roles, treatment stages, and resource constraints, while reducing repetitive or unrelated content. This approach aligns with broader challenges in clinical record summarization (Pivovarov & Elhadad, 2015).

The output of this stage is not a service plan. It is a structured input package for the six-step workflow, including the standardized case description, case planning profile, and list of missing or unclear information.

**5. Six-Step Planning Workflow**

The six-step planning workflow is the core procedural component of MedSWFlow and is the same workflow used in the companion expert evaluation study (Song et al., 2026). It divides medical social work case planning into six sequential stages: case assessment, problem analysis, goal setting, intervention planning, risk anticipation, and planned effect evaluation (Figure 2). Each stage receives a defined input, applies a specific framework, produces a structured output, and passes information to subsequent stages. The LLM generates content within each stage, while the workflow determines the planning sequence and output structure.

[Figure 2 about here]

The first three stages form the assessment section of the draft case plan, moving from understanding the case to identifying priority problems and goals. The last three stages form the service plan section, translating assessment findings into intervention modules, risk considerations, and evaluation indicators. The final output is a structured draft intended for practitioner review rather than a final professional decision. The six steps of the workflow, along with their inputs, frameworks, outputs, and constraints, are detailed below (Table 5).

**Table 5**

*Six-Step Planning Workflow in MedSWFlow*

| **Step** | **Input** | **Framework or rule** | **Output** | **Key constraint** |
|---|---|---|---|---|
| Case assessment | Standardized case input and case profile | ICF-informed assessment structure (World Health Organization, 2001) | Structured assessment across functional, participation, environmental, and personal domains | Organize available information; do not propose interventions or infer missing facts |
| Problem analysis | Case assessment output | COM-B and Behaviour Change Wheel logic (Michie et al., 2011) | One or two priority problems, barrier analysis, and possible intervention directions | Prioritize rather than list all problems; distinguish documented facts from assumptions |
| Goal setting | Problem analysis output | Rehabilitation goal-setting principles (Bovend'Eerdt et al., 2009; Wade, 2009) | Short-, medium-, and longer-term draft goals with observable criteria | Link goals to priority problems; avoid broad or unverifiable goals |
| Intervention | Assessment, | Intervention | Intervention | Specify |

| Step | Input | Framework or rule | Output | Key constraint |
|---|---|---|---|---|
| planning | problem analysis, and goal outputs | specification frameworks, including TIDieR and AACTT (Hoffmann et al., 2014; Presseau et al., 2019; Proctor et al., 2013) | modules with actions, actors, targets, context, intensity, and monitoring | responsibilities and conditions; avoid vague service recommendations |
| Risk anticipation | Outputs from Steps 1–4 | Implementation outcomes and unintended consequences frameworks (Proctor et al., 2011; Stratil et al., 2024) | Key risks, possible triggers, affected intervention modules, and priority levels | Identify risks; do not assume interventions are feasible or ethically adequate |
| Planned effect evaluation | Goal, intervention, and risk outputs | ICF-informed outcome review and implementation outcome concepts (Proctor et al., 2011; World Health Organization, 2001) | Draft outcome and process evaluation indicators | Align indicators with goals and interventions; avoid unverifiable indicators |

### 5.1 Case Assessment

The first stage organizes standardized case information into a structured assessment. MedSWFlow uses ICF-informed domains, including functioning, participation, environmental factors, and personal factors, to provide a consistent framework for organizing referral information (World Health Organization, 2001). The goal is not formal ICF coding but systematic case understanding.

The output summarizes health and functional status, participation limitations, contextual barriers and supports, and relevant personal factors. This assessment becomes the basis for later planning. The stage is limited to organizing available information and should not introduce interventions or unsupported assumptions.

### 5.2 Problem Analysis

The second stage identifies one or two priority problems and examines factors contributing to them. MedSWFlow applies the COM-B model and Behaviour Change Wheel, considering capability, opportunity, and motivation as potential influences on the problem (Michie et al., 2011). In medical social work, these factors may include client understanding, caregiver skills, family support, service access, trust, stigma, or readiness for change.

The output includes a priority problem statement, a brief explanation of contributing factors, and possible intervention directions. The emphasis is on prioritization rather than an exhaustive problem list.

**5.3 Goal Setting**

The third stage translates priority problems into draft service goals. Drawing on rehabilitation goal-setting principles, goals are expected to be specific, observable, and linked to a time frame (Bovend'Eerdt et al., 2009; Wade, 2009). Depending on the case, goals may be organized into short-, medium-, and longer-term objectives.

The output is a staged goal system that specifies the desired change, expected progress, and approximate timeline. Goals should remain closely aligned with the priority problems identified in the previous stage.

**5.4 Intervention Planning**

The fourth stage converts assessment findings and goals into intervention modules. MedSWFlow uses intervention specification frameworks to clarify what actions will be taken, by whom, for whom, under what conditions, and how implementation will be monitored (Hoffmann et al., 2014; Presseau et al., 2019; Proctor et al., 2013).

The output includes intervention activities and delivery details. The workflow emphasizes concrete actions rather than broad recommendations. Proposed interventions should not assume the availability of resources, personnel, or services unless supported by case information or explicitly marked for verification.

**5.5 Risk Anticipation**

The fifth stage identifies risks associated with the proposed intervention modules. Drawing on implementation outcome and unintended consequence frameworks, MedSWFlow considers issues such as service burden, role ambiguity, ethical concerns, poor contextual fit, resource limitations, and unintended negative effects (Proctor et al., 2011; Stratil et al., 2024).

The output is a structured list of risks linked to specific intervention modules, including possible triggers and priority levels. The purpose is to support professional review rather than to determine implementation readiness.

**5.6 Planned Effect Evaluation**

The final stage defines how progress may be evaluated. MedSWFlow combines ICF-informed outcome thinking with implementation outcome concepts to distinguish between outcome evaluation and process evaluation (Proctor et al., 2011; World Health Organization, 2001). Outcome evaluation focuses on changes in functioning, participation, support, or psychosocial status. Process evaluation examines whether planned activities were delivered and accepted as intended.

The output is a draft evaluation plan containing observable and verifiable indicators aligned with goals and interventions. Indicators may include completion of planned activities, engagement of family members, service utilization, or documented participation in decision-making.

Together, the six stages generate two draft documents: a case assessment form and a service plan. These outputs support practitioner review and revision. Human professionals remain responsible for verifying information, assessing feasibility, protecting client rights, and adapting plans to local service contexts.

## 6. Review Checkpoints, Safety, and Human Oversight

MedSWFlow generates planning drafts for a high-risk human service context. Its safety design emphasizes role boundaries, uncertainty management, and human oversight. The system does not provide medical diagnosis, treatment advice, medication advice, legal determinations, or final service decisions. All outputs require practitioner review before use.

### 6.1 Design Principles

Three principles guide the design. First, the LLM is limited to structured drafting within predefined planning stages. Second, uncertain or missing information should be marked rather than inferred. Third, responsibility for decisions remains with practitioners and supervisors. These principles align with recommendations for responsible AI use in health and human services (Perron, Goldkind, et al., 2025; Reamer, 2023; World Health Organization, 2021).

### 6.2 Prompt-Level Constraints

Constraints are embedded in stage-specific templates. Prompts restrict the system to medical social work planning support and prohibit diagnosis, treatment recommendations, medication guidance, legal judgments, or guarantees of outcomes. Missing information must be identified as unknown or requiring further assessment rather than assumed.

### 6.3 Workflow-Level Constraints

Each stage has defined limits. Assessment organizes case information without proposing interventions. Problem analysis identifies priorities without treating assumptions as facts. Intervention planning outlines possible actions while marking items that require verification. Risk anticipation highlights concerns rather than declaring a plan safe or feasible. These limits support review and revision.

### 6.4 Human-Review Checkpoints

Human review is required throughout the workflow. Practitioners should examine priorities,

feasibility, assumptions, ethical concerns, burden, and local resource availability (Song et al., 2026). Supervisory review is recommended for high-risk situations, including suicide, abuse, neglect, consent disputes, severe family conflict, privacy concerns, or medical instability. Key safety risks, the system's constraints, and the required human review actions are mapped to each other (Table 6).

**Table 6**

*Safety Risks, System Constraints, and Required Human Review*

| Risk area | System constraint | Required human review |
|---|---|---|
| Medical or legal overreach | No diagnosis, treatment advice, medication advice, or legal determination | Confirm role boundaries and refer when needed |
| Unsupported assumptions | Mark uncertain information as unknown | Verify facts, resources, and client preferences |
| Priority mismatch | Require priority identification before planning | Review urgency, safety, and service needs |
| Contextual infeasibility | Make resource and implementation assumptions explicit | Verify staffing, and resource availability |
| Excessive burden | Flag service, family, or coordination burden | Assess workload and caregiver capacity |
| Ethical or relational risk | Flag autonomy, consent, privacy, and boundary issues | Review client rights and professional responsibilities |
| High-risk situations | Drafts are not sufficient for action | Require supervisory review |

### 6.5 Traceability

Traceability in the current prototype is limited to the separation of case inputs, case profiles, prompt templates, stage outputs, and final documents. More systematic logging, prompt-version tracking, model metadata, and audit reports remain future implementation needs. MedSWFlow should not be assumed to provide enterprise-level audit trails or version tracking.

Overall, MedSWFlow does not make plans automatically safe, feasible, or ethically adequate. It makes assumptions and review requirements more visible. All outputs remain drafts that require human verification and revision.

## 7. Implementation

MedSWFlow is an open-source research prototype distributed as a local desktop application. The repository indicates that it is built with Python 3.13 and includes a Windows executable. Users run the system locally, configure model access, enter case information, and generate draft case plans through the workflow interface. The main components of the MedSWFlow codebase and their respective functions are listed below (Table 7).

**Table 7**

*Repository Structure of MedSWFlow*

| Component | Function |
|---|---|
| data/ | Stores case and prompt data |
| cases.json | Standard case library |
| prompts.json | Six-step workflow prompts |
| prompts0.json | Case standardization prompt |
| src/ | Core source code |
| run_pipeline.py | Main workflow entry point |
| GPT.py and GPT2.py | Model-call wrappers |
| demo.py, demo0.py, demo2.py | Demo and testing scripts |
| output/ | Generated workflow outputs |
| .env | Local model configuration |
| pyproject.toml and uv.lock | Project configuration and dependency locking |
| tests/ | Test directory, if populated in the current repository |

Cases are first standardized into the five-module structure and then processed through the six-step workflow. Prompt templates guide assessment, problem analysis, goal setting, intervention planning, risk anticipation, and planned effect evaluation. Outputs are combined into a draft case plan that can be exported as a Word document.

The system supports user-configured API access to LLMs, including ChatGPT and DeepSeek. Users provide API credentials and endpoint URLs, allowing the same workflow to operate across different models. MedSWFlow should be regarded as a reproducible research implementation rather than a production-ready practice system.

**8. Usage Cases**

This section presents three usage cases to illustrate how MedSWFlow behaves across different medical social work scenarios. The cases are based on standardized scenarios used to generate draft plans through the six-step workflow. They are not presented as evidence of effectiveness. They show how the workflow structures case information, identifies planning priorities, generates staged outputs, and leaves key decisions for practitioner review. Three illustrative cases and the key workflow behaviors they demonstrate are introduced below (Table 8).

**Table 8**

*Illustrative Usage Cases*

| Case | Planning challenge | Workflow behavior illustrated |
|---|---|---|
| High-risk adolescent psychiatric | Self-harm risk, family-system stress, treatment engagement, and social participation disruption | Priority setting, family-system formulation, and safety-oriented problem analysis |
| Older adult discharge and long-term care placement | Discharge delay, severe functional decline, family role conflict, cross-border family support, and care placement decision-making | Multi-party intervention planning, role clarification, and feasibility-sensitive service design |
| PICU acute bereavement | Sudden child death, acute parental crisis, administrative blockage, financial burden, and immediate safety concerns | Staged goal setting, crisis prioritization, and planned transition from acute stabilization to follow-up support |

### 8.1 High-Risk Adolescent Psychiatric Case

**Brief case background.** This case involves a 15-year-old student with depression, depersonalization symptoms, repeated wrist-cutting, and elevated suicide risk. The client had suspended school attendance. Family dynamics were a major concern. The father was frequently absent and tended to minimize the illness, while the mother provided intensive care but exercised strong control over daily activities and peer relationships. The psychiatrist requested social work involvement to assess family functioning, reduce conflict around recovery expectations, and explore whether the home environment supported treatment.

**Planning challenge.** The case requires attention to both individual risk and family-system factors. A narrow focus on emotional symptoms would overlook parental responses, social isolation, and disrupted developmental tasks. Recommendations about family communication would also be insufficient if immediate safety concerns were not addressed.

**Workflow behavior.** During problem analysis, MedSWFlow identified two priority areas. The first concerned self-harm and severe emotional distress. The generated formulation linked these behaviors to psychiatric symptoms, family pressure, perceived invalidation, and difficulties expressing autonomy. The second concerned loss of social participation, including school suspension, withdrawal from peers, and reduced engagement in daily routines. The draft connected these issues through interactions among emotional state, family environment, and available support.

**What the case illustrates and what requires review.** This case illustrates how the workflow organizes problem analysis before intervention planning. Immediate safety concerns appear before longer-term goals such as school reintegration or family education. However, the output remains a planning aid rather than a decision-making tool. Practitioners must verify suicide risk, treatment arrangements, guardian consent, family readiness, and institutional requirements. Decisions regarding safety planning, family intervention, or protective action require professional assessment and supervision.

### 8.2 Older Adult Discharge and Long-Term Care Placement Case

**Brief case background.** This case concerns a 92-year-old woman hospitalized after surgery, with delayed recovery, Parkinson's disease, severe functional decline, and resistance to some care activities. Hospital discharge had been delayed for more than four weeks. Her spouse and two sons lived overseas and mainly provided financial support. A daughter in Shanghai managed most communication with the hospital but was unwilling to assume full caregiving responsibility because of longstanding family tensions and caregiver burden. The referral focused on discharge planning, family role clarification, and transition to long-term care.

**Planning challenge.** The case involves multiple stakeholders with different responsibilities and expectations. Planning must consider the client's preferences, family relationships, caregiving capacity, institutional pressure, and available community resources. Broad recommendations such as "improve family communication" provide limited guidance unless responsibilities, timing, and practical constraints are specified.

**Workflow behavior.** During intervention planning, MedSWFlow generated two main modules. The first focused on supported decision-making with the client, including structured discussion of preferred living arrangements, care priorities, and acceptable levels of support. The second focused on family coordination and transition planning. The draft suggested a meeting involving the client, daughter, social worker, and potential receiving institution to clarify caregiving limits, financial responsibilities, monitoring arrangements, and placement conditions.

**What the case illustrates and what requires review.** This case illustrates how the workflow translates case information into explicit intervention components. Actions, actors, targets, timing, and monitoring considerations are specified rather than left as broad service categories. Feasibility remains uncertain until reviewed by practitioners. Bed availability, eligibility criteria, costs, decision-making capacity, family consent, and local discharge policies must all be verified. Professional judgment is required to determine whether proposed arrangements are realistic and appropriate.

### 8.3 PICU Acute Bereavement and Crisis Intervention Case

**Brief case background.** This case involves migrant parents whose eight-year-old child died within 24 hours of admission to the PICU for fulminant myocarditis. Following the death, the mother became largely unresponsive and stopped eating, while the father experienced intense guilt and emotional dysregulation. The family had limited local support, and relatives were unable to travel immediately because of the Spring Festival period. Unresolved medical expenses and administrative procedures created additional stress.

**Planning challenge.** The situation requires prioritization of acute crisis needs. The draft plan must address safety, communication, bereavement reactions, administrative barriers, and coordination with hospital procedures. It should neither reduce the situation to routine grief counseling nor focus only on paperwork without considering the parents' psychological state.

**Workflow behavior.** During goal setting, MedSWFlow generated staged objectives. Short-term goals focused on establishing a basic communication framework, reducing immediate behavioral risks, and restoring minimal participation in urgent decisions. Medium-term goals addressed completion of essential post-death procedures, involvement of relatives, and reduction of extreme self-blame or distrust. Longer-term goals focused on re-establishing daily functioning, connecting the family with psychosocial resources, and supporting ongoing adaptation to loss.

**What the case illustrates and what requires review.** This case illustrates how the workflow structures crisis planning through temporal sequencing. Immediate stabilization is addressed before administrative coordination, and longer-term adjustment is separated from acute crisis management. However, the output cannot determine decision-making capacity, legal obligations, institutional requirements, or the need for emergency psychiatric referral. Practitioners must assess risk, timing, consent, and service appropriateness in real time.

Across the three cases, MedSWFlow illustrates a consistent workflow behavior. It takes standardized case information, builds a planning profile, processes the profile through six stages, and produces structured draft content. The workflow can make planning elements more visible, but it cannot verify facts, determine feasibility, confirm consent, assess live risk, or replace professional judgment.

## 9. Discussion and Future Work

### 9.1 What MedSWFlow Does and Does Not Do

MedSWFlow is a workflow framework for drafting medical social work case plans. It organizes planning into a reproducible sequence covering assessment, problem analysis, goal setting, intervention planning, risk anticipation, and planned effect evaluation. By standardizing case input and applying staged prompts, it produces structured draft plans that are easier to inspect than outputs from a single free-form prompt.

MedSWFlow does not replace medical social workers, clinical teams, or supervisors. It does not make diagnoses, recommend treatment, determine legal responsibilities, assess eligibility, authorize services, or determine whether a proposed intervention is feasible or ethically appropriate in a specific context. Such decisions require professional judgment, local resource knowledge, client preferences, and institutional procedures.

### 9.2 Workflow-Guided Drafting Rather Than Autonomous Decision-Making

A key design choice in MedSWFlow is workflow-guided drafting rather than autonomous decision-making. The workflow defines the sequence of tasks and the constraints for each stage, while the connected LLM generates content within those boundaries. This approach supports interpretability, reproducibility, and professional review. Practitioners can identify which statements belong to assessment, goals, interventions, risks, or evaluation, and can more easily detect unsupported assumptions or unrealistic recommendations.

This design also has limits. Real-world medical social work is often iterative, with plans changing as new information emerges. A fixed workflow cannot fully capture this dynamic process. MedSWFlow should therefore be understood as a structured drafting environment rather than a complete model of professional reasoning.

### 9.3 Relationship to the Companion Expert Evaluation

The workflow described in this preprint is the same six-step workflow used in the companion expert evaluation study (Song et al., 2026). The companion study examined how expert medical social workers evaluated workflow-generated plans, while this preprint focuses on the workflow architecture, implementation, safety constraints, and illustrative use cases.

MedSWFlow was not redesigned after expert evaluation. The workflow and the evaluation study are parallel components of the same project. The companion study provides practitioner perspectives on generated drafts, whereas this preprint documents the underlying system. The evaluation findings reinforce the importance of treating outputs as drafts requiring human review, but they do not establish effectiveness in practice.

### 9.4 Current Limitations

Several limitations should be acknowledged. First, the examples presented are scenario-based and do not demonstrate real-world outcomes such as improved service quality, reduced workload, or better client results.

Second, MedSWFlow depends on prompt-following behavior. Although the workflow structures tasks and constraints, connected LLMs may still generate incomplete, vague, or unsupported content. The workflow supports transparency but cannot guarantee compliance.

Third, local resource integration remains limited. Case planning in practice depends on hospital procedures, community services, funding rules, referral pathways, staffing conditions, and time windows. Without institution-specific knowledge, generated recommendations may require careful verification.

Fourth, the current prototype provides only basic traceability. More comprehensive audit functions, including prompt version control, model logging, metadata retention, and structured audit reporting, require further development.

Finally, broader stakeholder evaluation is needed. Future assessments should include clients, caregivers, healthcare professionals, supervisors, and administrators, whose perspectives on usefulness, trust, privacy, and burden may differ from those of expert reviewers. Some situations, such as crises, mandated reporting, or psychiatric emergencies, require immediate professional action before any case-planning workflow can be applied.

### 9.5 Future Directions

Future work should focus on five areas. First, stronger integration with local resource databases, hospital procedures, and referral systems could make feasibility review more specific and reduce unsupported assumptions. Second, practitioner feedback loops could allow users to revise drafts, flag inaccuracies, and provide structured feedback for workflow refinement. Third, assumption detection could be strengthened by identifying recommendations that lack support from the case profile. Fourth, usability could be improved through stage-by-stage review panels, source-to-output comparisons, annotations, and supervisory dashboards. Fifth, prospective implementation studies are needed to examine how practitioners use MedSWFlow in practice-like settings. Relevant outcomes include documentation efficiency, perceived usefulness, review burden, plan quality, ethical concerns, and team communication.

In summary, MedSWFlow should be evaluated as a workflow for structured drafting rather than an autonomous practice system. Its value lies in making planning processes more transparent, producing reviewable drafts, and highlighting assumptions and risks. Safe use depends on professional judgment, local adaptation, resource verification, ethical oversight, and supervision.

## 10. Conclusion

MedSWFlow is an open-source, model-agnostic LLM workflow for drafting medical social work case plans. It structures case planning into six stages: assessment, problem analysis, goal setting, intervention planning, risk anticipation, and planned effect evaluation. By separating case input, profiling, prompting, and output, it makes the drafting process more transparent and reproducible.

The system generates reviewable drafts rather than final service plans. It can help practitioners organize case information, identify key problems, connect goals with interventions, and highlight issues requiring further review. MedSWFlow does not replace professional judgment, and all outputs should be verified by qualified practitioners.

As an open-source framework, MedSWFlow provides a basis for future research on workflow-guided LLM use in medical social work. Further studies should examine how practitioners revise generated drafts and whether such workflows can support documentation, review, and interdisciplinary coordination in practice.

## Appendix

**Figure 2.**Six-step flowchart

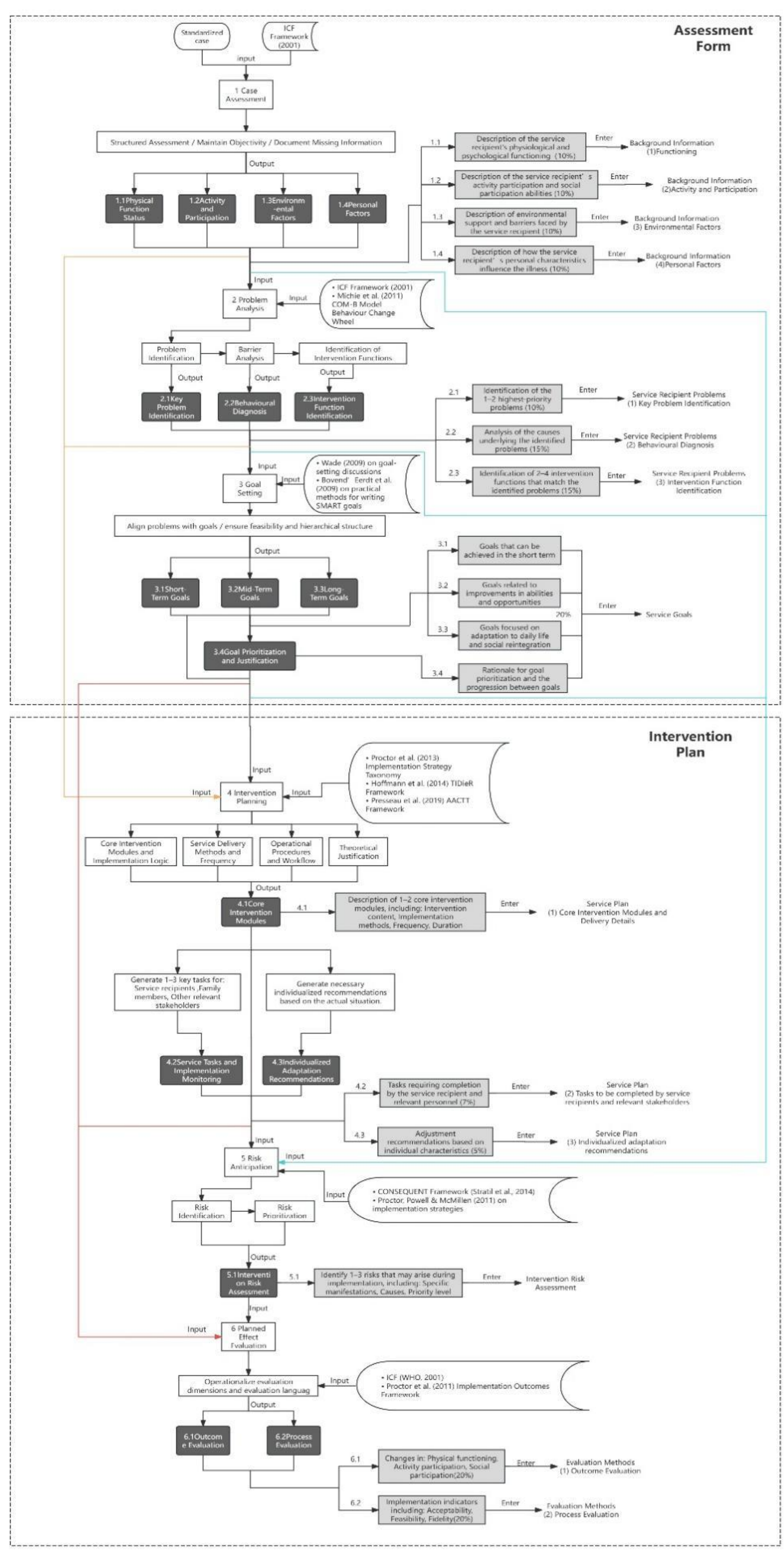